\documentclass[conference]{IEEEtran}
\IEEEoverridecommandlockouts

\usepackage{cite}
\usepackage{amsmath,amssymb,amsfonts}
\usepackage{algorithmic}
\usepackage{graphicx}
\usepackage{textcomp}
\usepackage{multirow}
\def\BibTeX{{\rm B\kern-.05em{\sc i\kern-.025em b}\kern-.08em
    T\kern-.1667em\lower.7ex\hbox{E}\kern-.125emX}}

\usepackage{psfrag} 
\usepackage[hidelinks]{hyperref}
\usepackage[dvipsnames]{xcolor}  
\usepackage{pgfplots}       
\pgfplotsset{compat=newest} 
\pgfplotsset{plot coordinates/math parser=false} 
\usetikzlibrary{plotmarks}  
\usepackage{import}         

\begin{document}

\title{Cross-Comparison of Neural Architectures and\\ Data Sets for Digital Self-Interference Modeling\\[-0.5ex]
\thanks{This work is funded by the German Research Foundation (DFG) under Grant EN 869/4-1 with Project ID 449601577.}
}

\author{
\IEEEauthorblockN{Gerald Enzner, Niklas Knaepper, Aleksej Chinaev
\IEEEauthorblockA{\textit{Department of Medical Physics and Acoustics},
\textit{Carl von Ossietzky Universität Oldenburg},
26111 Oldenburg, Germany \\
\{gerald.enzner, niklas.knaepper, aleksej.chinaev\}@uni-oldenburg.de}
\vspace{-3ex}
}
}

\maketitle

\begin{abstract}
Inband full-duplex communication requires accurate modeling and cancellation of self-interference, specifically in the digital domain.  Neural networks are presently candidate models for capturing nonlinearity of the self-interference path. This work utilizes synthetic and real data from different sources to evaluate and cross-compare performances of previously proposed neural self-interference models from different sources. The relevance of the analysis consists in the mutual assessment of methods on data they were not specifically designed for. We find that our previously proposed Hammerstein model represents the range of data sets well, while being significantly smaller in terms of the number of parameters. A new Wiener-Hammerstein model further enhances the generalization performance.
\end{abstract}
\vspace*{1ex}
\begin{IEEEkeywords}
system modeling, neural networks, full-duplex
\end{IEEEkeywords}

\section{Introduction}
\label{sec:intro}
Neural networks in communications may refer to channel modeling, receiver design, and detection \cite{IEEE_BITS_overviewCommDNN}. A particular field of interest consists in the modeling of self-interference channels for inband full-duplex (IBFD) systems. IBFD relies on the same frequency resource for simultaneous transmission and receiving and, in this way, can cope with limited availability of bandwidth and maintain the ubiquitous requirement of low latency \cite{Spectrum2021}. However, as a matter of an extremely low signal-to-self-interference ratio, an overall self-interference cancellation (SIC) in the order of 100\,dB would be required for successful system operation \cite{Wichmann2014,Smida2024}. Pioneering work therefore joins the forces of active or passive SI shielding in the propagation domain, adaptive or non-adaptive cancellation in the analog receiver unit, and adaptive cancellation in the digital baseband section of the receiver \cite{Heino_2015,Herd_2019,Smida2023}.

For the sake of distortionless response regarding the signal of interest, when processing the received signal, the majority of SIC systems relies on the idea of subtractive cancellation of the self-interference. This operation paradigm is thus akin to concepts of nonlinear system identification, such that a duplicate of the nonlinear self-interference path must be determined with very accurate requirements regarding the prediction of the self-interference (SI) based on the known transmission signal. When the desired system identification is approached with neural network representations, their concept of nonlinear activation naturally embodies the inevitable nonlinearity of the RF self-interference path by generic computational modeling \cite{PANSE2022101526}. Specific networks have been shown to improve over traditional polynomial modeling accuracy, while reducing computational complexity \cite{Zhang_2018,Guo2019,Stimming2019, bilbao2025cnn}. Considering the potential challenge of importing neural network models to realworld embedded systems, even realtime assessment of neural networks was already demonstrated in~\cite{Kurzo2020}. Frequently, neural SI modeling relies on a two-stage decomposition model of firstly linear and secondly nonlinear SIC in order to focus the optimization of the nonlinear representation on the truly nonlinear part of SI signals   \cite{Stimming2019,Baek2019,Muranov2021}. Most of the available systems share the idea of a very interpretable architectural setup based on underlying physical domain knowledge.  

Despite the trend of neural network modeling and related investigations, we shall still reflect the history of SIC based on sophisticated principles of signal processing and estimation, for instance, using maximum-likelihood \cite{Le-Ngoc_2016}, subspace \cite{Le-Ngoc_2017}, mean-square error \cite{Tepedelenlioglu_2018}, least-squares \cite{Guo_2023}, linear \cite{Zhou_2023} and/or polynomial adaptive-filter methods \cite{Kiayani2018,Vogt_2019}. Additionally, the design of robust receivers via non-convex optimization \cite{Sezgin2020} or modified matched filtering \cite{Mohammad_2023} was reported. The transmitter side as well can be optimized by active injection of compensation signal \cite{Le-NgocTho2017FWCS}, by the choice of pilot sequences with favorable nonlinear behavior \cite{Kong_2019}, or by the nonlinear predistortion of RF transmitters \cite{Austin2017,Gregorio2017}.

Returning to neural modeling for nonlinear system identification, a delicate factor of the optimization consists in the choice of the training and evaluation data. The authors of  \cite{Stimming2019,balatsoukas2018non}, for instance, have investigated feedforward and recurrent neural networks with memory using signals recorded from their testbed in \cite{Balatsoukas_2015}. Our own prior work \cite{enzner2024neural} has constructed synthetic data according to a memoryless power amplifier nonlinearity followed by an indoor wave propagation model \cite{Erceg_2004,Chen_2018}. The data of this block-structured Hammerstein system was successfully represented with a related neural architecture. However, depending on the particular data, the related architectural designs and training may overfit, such that network utility may be restricted to the data at hand. This paper therefore makes three additional contributions: 
\begin{itemize}
    \item cross-comparison of data and networks from the previous studies in \cite{Stimming2019} and \cite{enzner2024neural} in terms of SI modeling,\vspace*{0.25ex}
    \item derivation of an extended network architecture in order to represent the data of \cite{Stimming2019} and \cite{enzner2024neural}, and\vspace*{0.25ex}
    \item evaluation of the linear/nonlinear decomposition model \cite{Stimming2019} against monolithic nonlinear representation \cite{enzner2024neural}.
\end{itemize}
The remainder of the paper is organized as follows. Section~\ref{sec:system} describes the different data sets, Section~\ref{sec:architectures} refers to different neural network models, and Section~\ref{sec:results} delivers the results.

\section{Self-Interference Data}
\label{sec:system}

This section reflects on the diversity of the different data sets used in this study. We refer to synthetic data in two configurations (Hammerstein and Wiener structured) and to a set of real recordings from an independent source. 

\subsection{Synthetic Hammerstein Data}

A simulated full-duplex transceiver after \cite{enzner2024neural,Duarte_2012,Valkama_2018} is shown in Fig.~\ref{fig:baseline}~(a). Our generation of the SI signal therein is based on the baseband discrete-time transmission signal~$s[k]$.~It passes digital-to-analog (D/A) conversion, nonlinear power amplification (PA), and transmission over a wireless linear channel $h_{\text{SI}}(t)$ to the receiver, together referred to as a ''Hammerstein'' arrangement (generally consisting of memoryless nonlinearity followed by a linear system). With additional receiver noise, the input-output relation between $s[k]$ and a discrete-time version $y_H[k]$ of the analog SI $y_H(t)$ is thus
\begin{equation}
    y_H[k] = \text{PA} \big( s[k] \big) * h_{\text{SI}}[k] + n_H[k] \;,
\end{equation}
\begin{equation}
    \text{PA} \big( s\big) = F\big( |s|\big) \cdot e^{j\cdot \text{arg}(s)}, \; \; F\big( |s|\big) = f \cdot \text{arctan}(c_f \cdot |s|) \;,
    \label{eq:PA}
\end{equation}
such that the PA assumes only amplitude-to-amplitude distortion \cite{Eltawil_2015, Joung_2014}, where $f$ and $c_f$ are parameters controlling the output power and degree of nonlinearity, respectively. The linear SI channel relies on the WLAN multipath fading 'Model C' \cite{Erceg_2004} and is implemented by means of the Matlab WLAN Toolbox~\cite{Cho_2010}. For illustration, the PSDs of the SI signal and the receiver noise floor are shown in Fig.~\ref{fig:psdOverview} (top). 

The actual input-output characterization of the SI path in a transceiver varies by the point of SIC, which is in Fig.~\ref{fig:baseline}~(a) located in the analog domain just before the LNA, and by the (neural network) SIC model input, which here is the transmission signal in the digital domain. The related Hammerstein characterization of the SI path has been represented with a matched Hammerstein network architecture for SIC in \cite{enzner2024neural}.

\subsection{Synthetic Wiener Data}

\begin{figure}[!tb]
  \centering 
\psfrag{SI}[][]{\footnotesize{SI}}
\psfrag{TX}[][]{\footnotesize{TX}}
\psfrag{RX}[][]{\footnotesize{RX}}
\psfrag{SIC}[][]{\footnotesize{SIC}}
\psfrag{PA}[][]{\footnotesize{PA}}
\psfrag{LNA}[][]{\footnotesize{LNA}}
\psfrag{D/A}[][]{\footnotesize{D/A}}
\psfrag{A/D}[][]{\footnotesize{A/D}}
\psfrag{ATT}[][]{\footnotesize{ATT}} 
  \psfrag{f(s)}[][]{\small $\!f(s)$}
  \psfrag{g(x)}[][]{\small $g(x)$}
  \psfrag{w(k)}[][]{$w_k$}
  \psfrag{s(k)}[][]{$s[k]$}
  \psfrag{r(k)}[][]{$r[k]$}
  \psfrag{y(k)}[][]{$y_\text{W}[k]$}
  \psfrag{z(t)}[][]{$z(t)$}
  \psfrag{hSI}[l][l]{\,$h_\text{SI}(t)$}
  \psfrag{yH}[][]{$\!\!\!y_\text{H}(t)$}
  \psfrag{x(t)}[][]{$x(t)$}
  \psfrag{adaptive system}[l][l]{\small \,\,\,\,neural net}
  \includegraphics[width=\columnwidth]{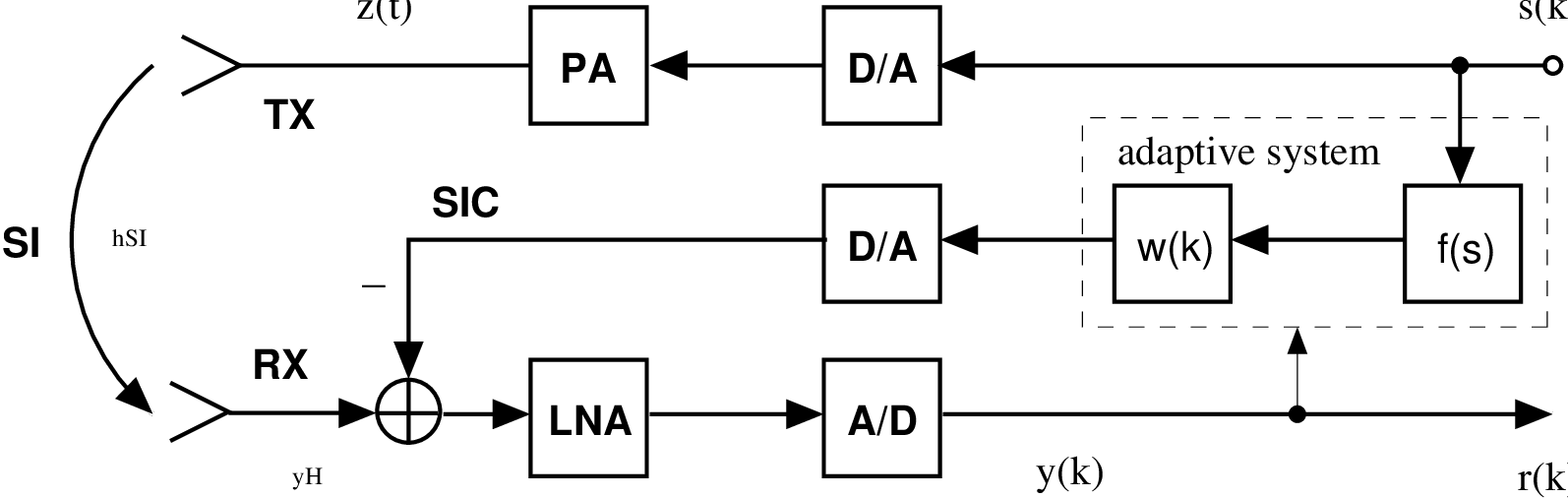}\\[1ex]
  (a) SIC system option characterizing ''Hammerstein'' data\\[4ex]
  \includegraphics[width=\columnwidth]{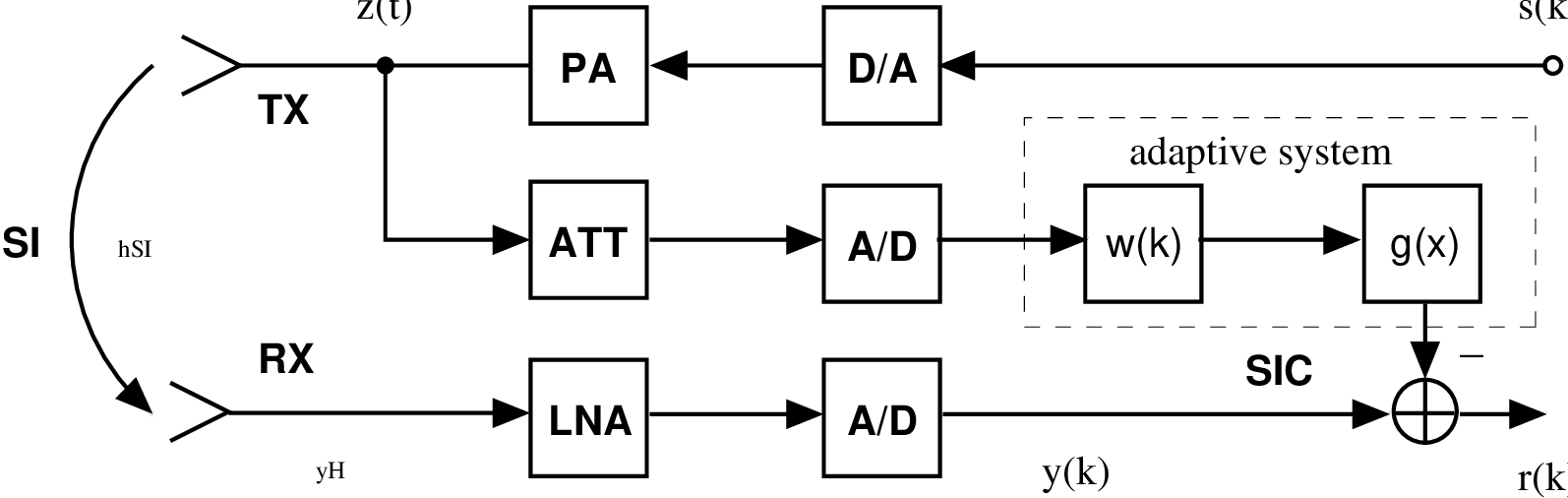}\\[1ex]
  (b) SIC system option characterizing ''Wiener'' data\\[0ex]
  \caption{System options with self-interference cancellation \cite{enzner2024neural}.}
  \label{fig:baseline}
\end{figure}

\begin{figure}
    \centering
    \includegraphics[width=0.97\columnwidth]{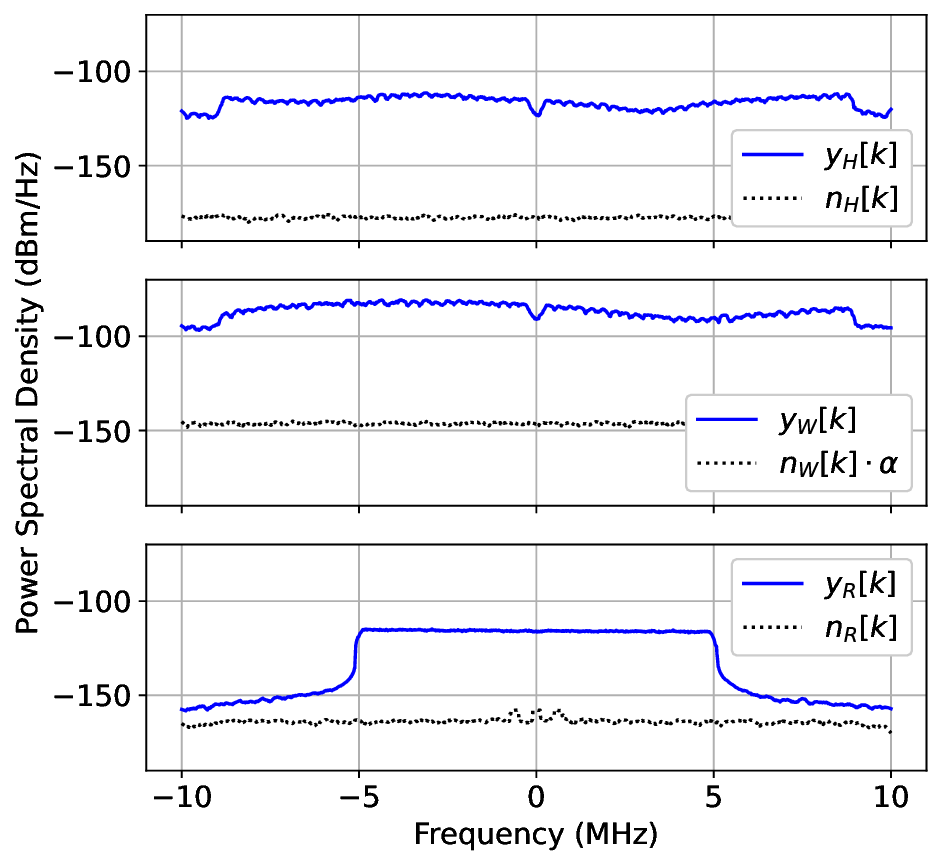}
    \vspace*{-2ex}
    \caption{The power spectral densities (PSDs) of SI signals}
    \vspace{-1ex}
    \label{fig:psdOverview}
\end{figure}

Considering the SIC arrangement of Fig.~\ref{fig:baseline}~(b), we have a (neural network) SIC model input $z(t)$ in the analog domain and our point of SIC is defined by the digital SI signal $y_W[k]$. The signal $z(t)$ here firstly passes the linear SI channel and then a low-noise-amplifier (LNA) and a saturating analog-to-digital (A/D) converter \cite{enzner2024neural,Korpi_2014, Eltawil_2015}, together referred to as a ''Wiener'' arrangement (generally a linear system followed by nonlinearity). With additional receiver noise, the SI reads
\begin{equation}
    y_W[k] = \text{AD} \bigg( \big( z[k] * h_{\text{SI}}[k] + n_W[k] \big) \cdot \alpha \bigg)\;,
\end{equation}
\begin{equation}
    \!\!\text{AD}\big( y\big) = G\big( |y|\big) \cdot e^{j\cdot \text{arg}(y)}, \; \; G\big( |y|\big) = \begin{cases}
    |y|, \; |y| < c_g \\
    c_g, \;\; |y| \geq c_g\;,\!\!
    \end{cases}
    \label{eq:AD}
\end{equation}
where $\alpha$ is the LNA gain of the simulation and AD is our nonlinear saturation function with threshold $c_g$. The PSDs of SI and noise floor including LNA amplification are again shown in Fig.~\ref{fig:psdOverview} (middle). The input-output data is available\footnote{\href{https://github.com/STHLabUOL/CrossCompareSICforIBFD}{https://github.com/STHLabUOL/CrossCompareSICforIBFD}} for Hammerstein and Wiener configurations. Since our study is mainly concerned with system identification, we only briefly mention the QPSK-based OFDM modulation of the transmit signal $s[k]$ in the complex baseband according to IEEE-802.11~\cite{Qureshi_2023} for the n-channel with $20\,\text{MHz}$ bandwidth \cite{enzner2024neural}.

\subsection{Real Recordings of SI Signals}
\label{subsec:real}

As a counterpart to the synthetic data, we further consider real WLAN signals, recorded and made available\footnote{\href{https://paperswithcode.com/paper/non-linear-digital-self-interference}{https://paperswithcode.com/paper/non-linear-digital-self-interference}} by the authors of \cite{Balatsoukas_2015,Stimming2019,balatsoukas2018non}. To capture these recordings, a National Instruments FlexRIO PXIe-1082 chassis and two FlexRIO 5791R RF transceiver modules were used. Like with our synthetic data, a QPSK modulation scheme was~configured in this system, but with a passband of only 10 MHz bandwidth. More detailed information regarding the wireless channel or the recording environment is not explicitly available.

For illustration, Fig.~\ref{fig:psdOverview} (bottom) compares the PSDs of SI and noise floor of the real recordings, denoted by index "R", with the former synthetic data. The biggest difference seems to be given by the smaller bandwidth of the real signals. However, we also observe a largely flat PSD of the real signals within its occupied bandwidth, which suggests a wireless SI channel with a pronounced ''Dirac'' shape and, thus, a dominant line of sight between transmitter and receiver.

\section{Neural-Network Architectures}
\label{sec:architectures}

This section reflects diverse model architectures with the idea of system identification. Two architectures are akin to our two types of synthetic data. A combined model thereof is proposed and another is used from an independent source.  

\subsection{Hammerstein Architecture}

We firstly refer to the ''Global Hammerstein'' architecture from \cite{enzner2024neural}, which structurally fits the Hammerstein logic of  Fig.~\ref{fig:baseline}~(a) for complex-valued computational modeling in the SIC cancellation path. Specifically, it comprises a nonlinear multi-layer perceptron (MLP) to model the nonlinear PA in the magnitude domain, followed by a purely linear convolution layer (Conv1D) to represent the linear transmission channel $h_{\text{SI}}(t)$. This model is depicted high-level in Fig.~\ref{fig:modelsOverivew}~(a).

\subsection{Wiener Architecture}

Similarly, a Wiener model architecture can be devised in order to structurally align with the opposite Wiener logic of Fig.~\ref{fig:baseline}~(b), i.e., switching nonlinear MLP and linear convolutional stages of the network as shown high-level by  Fig.~\ref{fig:modelsOverivew}~(b), where the convolution still represents the linear SI channel $h_{\text{SI}}(t)$ and the MLP models LNA and A/D saturation. 

\begin{figure}[!tb]
  \centering 
  \psfrag{Conv2D,1}[c][c]{Conv2D(1,1), n=8}
  \psfrag{Conv2D,2}[c][c]{Conv2D(1,1), n=1}
  \psfrag{Activation}[c][c]{Activation}
  \psfrag{complex}[c][c]{complex}
  \psfrag{mag.}{magnitude}
  \psfrag{phase}{phase}
  \includegraphics[width=0.45\columnwidth]{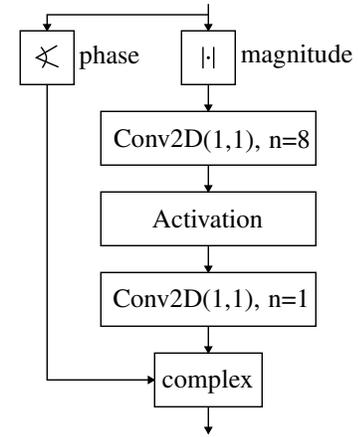}
  \caption{Magnitude-to-magnitude MLP with complex-valued input/output.}
  \vspace{-1ex}
  \label{fig:MLP}
\end{figure}

The MLP stage of both architectures is depicted in Fig.~\ref{fig:MLP} in more detail. It consists of one (or possibly more) hidden layers with, for instance, n=8 non-linear units and is followed by a linear output layer (i.e., suitable for the regression problem) with a single unit. The trainable weights are implemented as ''Conv2D''  layers for sequence processing, using kernels of size 1x1, thus acting independently on each sample. In accordance with Eqs.~(\ref{eq:PA}) and (\ref{eq:AD}), this MLP nonlinear function is exclusively applied to input magnitudes. As indicated by Fig.~\ref{fig:modelsOverivew}, a ''tanh'' activation is employed in the Hammerstein model to capture soft PA nonlinearity, while ReLU activation is preferred in the Wiener model for better fit with the synthetic clipping nonlinearity of our A/D simulation.

\subsection{Wiener-Hammerstein Architecture}

Both block-structured architectures may also be united into a more comprehensive Wiener-Hammerstein model for generalized representation of synthetic Wiener or Hammerstein data with a single model. As shown by Fig.~\ref{fig:modelsOverivew}~(c), the Wiener-Hammerstein model shares the convolutional layer with both the Wiener and Hammerstein models, and respective MLPs with either the Wiener or the Hammerstein model. While Wiener and Hammerstein models are supposed to specifically mimic the mappings from $s[k]$ to $y_H[k]$ and from $z[k]$ to $y_W[k]$ according to Figs.~\ref{fig:baseline}~(a) and (b), the Wiener-Hammerstein model may be applied to learn either of the two functions (hence the alternative I/Os  in the high-level diagram).

\subsection{Time-Delay Feedforward Neural Network (FFNN)}

\begin{figure}[!tb]
    \psfrag{MLP1}{MLP(tanh)}
    \psfrag{MLP2}{MLP(ReLU)}
    \psfrag{Conv1D}{Conv1D}
    \psfrag{FFNN}{FFNN}
    \psfrag{(a)}{(a) Hammerstein}
    \psfrag{(b)}{(b) Wiener}
    \psfrag{(c)}{\parbox{3cm}{(c) Wiener-\\Hammerstein}}
    \psfrag{(d)}{\parbox{3cm}{(d) Time-delay\\ \hspace*{.420cm} FFNN}}

    \psfrag{s[k]}{$s[k]$}
    \psfrag{y[k]}{$\widehat{y}_r[k]$}
    \psfrag{s[k]...}{\{$s[k], s[k-1],..$\}}
    \psfrag{z[k]}{$z[k]$}
    \psfrag{yH[k]}{$\widehat{y}_H[k]$}
    \psfrag{yW[k]}{$\widehat{y}_W[k]$}
    \psfrag{s[k],z[k]}{$s[k]$ or $z[k]$}
    \psfrag{yH[k],yW[k]}{$\widehat{y}_H[k]$ or $\widehat{y}_W[k]$}
  \centering
  \includegraphics[width=.98\columnwidth]{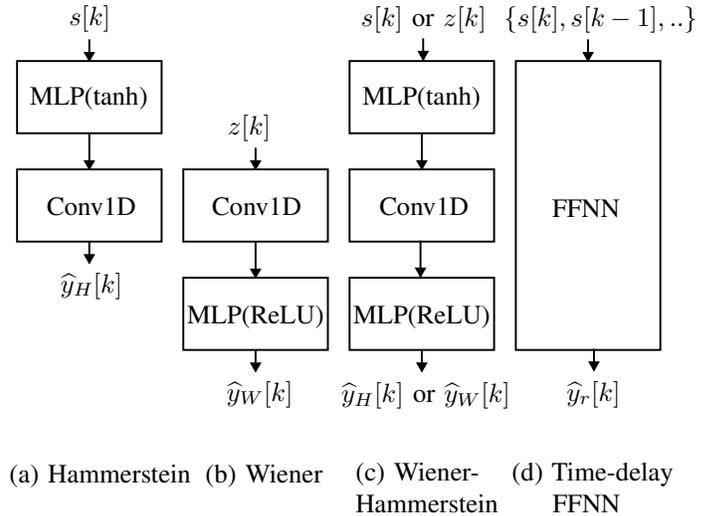}
  \vspace{1ex}
  \caption{Overview of diverse model architectures under test}
  \vspace{1ex}
  \label{fig:modelsOverivew}
\end{figure}

As a counterpart to these block-structured models, we further compare the monolithic, possibly more generic neural SI model from \cite{Stimming2019,balatsoukas2018non}, which not resembles either of the particular synthetic data generations, but was successfully evaluated on the real data (cf.~Sec.~\ref{subsec:real}) recorded by the authors. According to Fig.~\ref{fig:modelsOverivew}~(d), this FFNN architecture fully connects to a complex-valued temporal input context in order to predict the SI by means of two hidden layers with 26~and~17 units (using ReLU activation) and a linear output layer.

The FFNN system was originally used with linear premodeling, i.e., for two-stage linear and nonlinear SIC, and we shall explore such procedure with all architectures of Fig.~\ref{fig:modelsOverivew}. 

\begin{table*}[h] 
\begin{center}
\bgroup
\def\arraystretch{1.15}%
\caption{Neural Network Effort and Digital SIC performance}
\label{table:SICoverview}
\begin{tabular}{| l | c | c | c | c | c |}
\cline{4-6}
\multicolumn{3}{c}{~} & \multicolumn{3}{|c|}{\textbf{Self-Interference Attenuation [dB] on Different Data Sets}} \\
\hline
 \textbf{Model-Architectures} & \textbf{\#Parameters} & \textbf{GMACs} & \textbf{\quad Real Recordings\quad} & \textbf{Hammerstein Data} & \textbf{\qquad Wiener Data\qquad\qquad}\\  
 \hline
 Linear Model & $2\cdot13$ & 0.00053 & 37.9 & 10.2 & 10.5\\  
 \hline
 Time-Delay FFNN (+linear premodeling)  & 495\,\, (+$2\cdot13$) &  0.01468 & \textbf{39.8} \quad (\textbf{44.6}) & 20.2 \quad (21.1) & 30.3 \quad (31.0) \\  
 \hline
 Hammerstein (+linear premodeling) & 42\,\, (+$2\cdot13$) & 0.00143 & 38.1 \quad (42.9) & \textbf{53.3} \quad (39.3) & 12.0 \quad (12.6) \\  
 \hline
 Wiener (+linear premodeling)  & 51\,\, (+$2\cdot13$) & 0.00152 &  39.0 \quad (42.4)& 14.3 \quad (10.3) & \textbf{59.9} \quad (40.6)\\  
 \hline
 Wiener-Hammerstein (+linear premodeling) & 66\,\, (+$2\cdot13$)  &  0.00187 &  39.1 \quad (42.9) & \textbf{58.1} \quad (39.6) & \textbf{59.7} \quad (12.6) \\  
 \hline
\end{tabular}
\egroup
\end{center}
\vspace{-3ex}
\end{table*}

\section{Training and Evaluation}
\vspace{-0.25ex}
\label{sec:results}

This section implements the cross-comparison of all network models (with optional linear premodeling in all cases) against all synthetic and real data sets as described before. The amount of  synthetic data generation \cite{enzner2024neural} matches the given signal length of the real recordings of 20.000 discrete-time samples \cite{balatsoukas2018non}. These signals are split 90/10 into training/test data and we employ the Adam optimizer with MSE loss in all cases. Training is terminated individually after suitable epochs for the convergence of the different architectures.

Considering \cite{balatsoukas2018non,Stimming2019}, and \cite{enzner2024neural}, we configure a temporal context of 13 discrete-time samples in the complex-valued Conv1D layers (i.e., kernel-size) of all block-structured models and also in the FFNN input layer to represent the length of the linear transmission channel in each case. Further considering $8+8=16$ trainable weights of the MLP (no biases), the Hammerstein model amounts to $16+2\cdot13=42$ parameters, the Wiener model (using biases) to 50 parameters, and the Wiener-Hammerstein model to 66 parameters overall. In contrast, the fully-connected FFNN uses 495 parameters, which is one order of magnitude larger than the block-structured models. Table~\ref{table:SICoverview} summarizes the model sizes and the related Giga Multiply-Accumulate operations (GMACs) for each network architecture next to respective SIC performances in terms of self-interference attenuation (SIA) for different data sets.

When dealing with the real data, the FFNN model appears slightly superior to block-structured Hammerstein, Wiener, and Wiener-Hammerstein models, considering good 39\,dB against 38\,dB SIA without, and good 44\,dB against 42\,dB with a linear premodeling stage. Interestingly, a linear model alone already yields almost 38\,dB SIA, which indicates mostly linear behavior of the real data. With the model performances by and large comparable, a criterion for model selection might be the clearly lower complexity of block-structured models.

Turning our attention to synthetic Hammerstein and Wiener data for which we know of its stronger nonlinearity \cite{enzner2024neural}, it is not surprising that the SIA performance of the purely linear model deteriorates very much to only 10\,dB. Based on this observation, we further anticipate that linear premodeling will also not support overall SIA performance, which is confirmed by all SIA numbers in parenthesis in Table~\ref{table:SICoverview}. 

Obvioulsy, the Hammerstein architecture delivers very good SIA performance of about {53\,dB for the Hammerstein data, while the Wiener architecture similarly matches to the Wiener data as shown by 
good 59\,dB SIA. Contrary to this close correspondence of model and data, expectedly, the Hammerstein and the Wiener architectures mutually fail on flipped (i.e., mismatched) Wiener and Hammerstein data sets. However, we can nicely show that the proposed generalization in terms of the Wiener-Hammerstein architecture successfully manages to cope with both the specific Wiener \emph{and} Hammerstein data by attaining 
58\,dB and 59\,dB SIA in these two cases.

Given the successful combination of the Hammerstein and Wiener models into a composite Wiener-Hammerstein model with greater generalization capabilities, we return to a comparison with the FFNN model. We again highlight that, despite the increased model size of Wiener-Hammerstein (WH) against indidividual Wiener and Hammerstein models, the WH model is still one order of magnitude smaller than the FFNN model. Moroever, the aforementioned SIA of the WH model is clearly superior to only 20\,dB and 
30\,dB SIA of the FFNN model for synthetic Hammerstein and Wiener data sets, respectively. The FFNN limitation in these cases can be attributed to the lack of a convolutional layer which is dedicated (as in Wiener and Hammerstein models) to representing the pronounced linear SI channel of the synthetic data.

In a grand comparison of proposed block-structured Wiener-Hammerstein models against the FFNN representation, we naturally appreciate the highest 44\,dB SIA performance of FFNN (with linear premodeling) in the domain of the real recordings, which is gold standard. On the other hand, the 
block-structured models with about 42\,dB are not
far behind in this case. If the real signals would have passed a more pronounced indoor linear SI channel (so that the spectrum in Fig.~\ref{fig:psdOverview} would be considerably shaped also) we would expect some FFNN deterioriation as shown for synthetic data. The Wiener-Hammerstein model is in that respect well prepared for more challenging real recordings. However, we noticed that the optimization of the Wiener-Hammerstein model at times has been more tedious, presumably due to local minima.

\section{Conclusions}

This paper investigated a range of neural network models and data sets available for self-interference modeling. In this context, we have also contributed a low-complexity composite Wiener-Hammerstein model with stronger generalization than individual Wiener and Hammerstein models and with good applicability to real data. Additionally, the investigation showed that the acquisition of real data requires further attention, for instance, in terms of more pronounced nonlinearity and linear SI channel, for conclusive competition of network models. 

\bibliographystyle{IEEEbib}
\bibliography{refs}

\end{document}